# Quantum Probabilities Are Objective Degrees of Epistemic Justification

Philipp Berghofer

University of Graz

philipp.berghofer@uni-graz.at

**Abstract**

QBism is currently one of the most widely discussed "subjective" interpretations of quantum mechanics. Its key move is to say that quantum probabilities are personalist Bayesian probabilities and that the quantum state represents *subjective degrees of belief*. Even probability-one predictions are considered subjective assignments expressing the agent's highest possible degree of certainty about what they will experience next. For most philosophers and physicists this means that QBism is simply too subjective. Even those who agree with QBism that the wave function should not be reified and that we should look for alternatives to standard $\psi$-ontic interpretations often argue that QBism must be abandoned because it detaches science from objectivity. The problem is that from the QBist perspective it is hard to see how objectivity could enter science. In this paper, I introduce and motivate an interpretation of quantum mechanics that takes QBism as a starting point, is consistent with all its virtues, but allows objectivity to enter from the get-go. This is the view that quantum probabilities should be understood as objective degrees of epistemic justification.

## 1. Introducing QBism

Contemporary philosophy of quantum mechanics is dominated by so-called "quantum theories without observers" (Dürr & Lazarovici 2020, viii; Goldstein 1998). These are interpretations that seek to deliver a purely objective third-person perspective that does not contain any irreducibly subjective/operational concepts or perspectival moments. The dominant view among philosophers is that physicists teach quantum mechanics in a way that is not only misleading but borderline unscientific. This is because the concept of "measurement" plays a central role in textbook quantum mechanics. As Maudlin criticizes: "A precisely defined physical theory […] would never use terms like 'observation,' 'measurement,' 'system,' or 'apparatus' in its fundamental postulates. It would instead say precisely *what exists and how it behaves*" (Maudlin 2019, 5). Instead, proponents of objectivist[1] interpretations insist that "the goal of physics must be to formulate theories that are so clear and precise that any form of interpretation […] is superfluous" (Dürr & Lazarovici 2020, viii). The most prominent interpretations of that kind are Bohmian mechanics, the many-worlds interpretation, and GRW.

Each of these interpretations is in tension with some of our common-sense intuitions and struggles with conceptual problems. Regarding our intuitions: Bohmian mechanics and GRW are modificatory interpretations. They arguably have the virtue of getting rid of the notion of measurement but they do so by *modifying* the formalism of quantum mechanics. However, according to our common-sense approach to scientific practice, we would typically say that we should modify a successful scientific theory only if the new theory is empirically more successful. But here the opposite is the case. While quantum mechanics is the empirically most successful scientific theory in the history of science, Bohmian mechanics and GRW are less successful in their predictive power because in contrast to standard quantum mechanics we don't have a relativistic extensions for these theories (see Wallace 2023). In other words, many scientists share the intuition that if you have to choose between a highly successful scientific theory and a modification of it that is less developed and less successful in its predictive power, you should choose the former even if it has unexpected philosophical-ontological consequences. It is therefore no surprise that while Bohmian mechanics in particular is popular in philosophy, it remains widely ignored by physicists. The many-worlds interpretation, by contrast, has counter-intuitive consequences in the sense that it violates the principle of ontological parsimony as well as the idea that science should not postulate entities that are in principle unobservable.

One of the main conceptual problems of all "quantum theories without observers" is the question of how to make sense of the wave function. In textbook quantum mechanics, as well as in the many-worlds interpretation and in GRW, the state of the system is completely described by the wave function. (In Bohmian mechanics, the state is described by the wave function plus the positions of the particles.) The most natural move from the standard realist perspective would be to say that we should reify the wave function, i.e., consider it to be physically real. The analogous move made a lot of sense in classical mechanics. In the Newtonian picture, the state of the system is completely described by the positions and momenta of point particles and it seemed straightforward to assume that these quantities represent physically real objects/properties. But while the particles in classical mechanics are defined in three-dimensional Euclidean space, the wave function is defined in infinite-dimensional Hilbert space. Reifying the wave function thus means reifying the mathematical space it is defined in and while several authors argued for a version of configuration space realism (e.g. Albert 1996, 277) or indeed Hilbert space realism (Carroll & Singh 2019), other prominent voices in the "objectivist" camp have explicitly argued against any such version of wave function realism (Wallace 2021).

---

1   I use "objectivist interpretations" synonymously with "quantum theories without observers," their characteristic feature being that they seek to provide a purely objective third-person description of how external objects evolve in time.

What is more, most physicists accept the "statistical interpretation" of Born according to which the wave function is to be understood as a probability amplitude that has the mere instrumental function of telling us the probability that some quantum measurement will yield a given result. This is to say that physicists typically deny that the wave function is physically real and accordingly that the "collapse" of the wave function should be understood as a physical process. Insisting, by contrast, that the wave function is real leads to the measurement problem in the following sense: Why is it the case that the evolution of the wave function is completely determined by the Schrödinger equation, unless a measurement takes place? Even strong opponents of objectivist interpretations concede that subjective interpretations can easily avoid the measurement problem: "Any approach according to which the wave function is not something real, but represents a subjective information, explains the collapse at quantum measurement perfectly: it is just a process of updating the information the observer has" (Vaidman 2014, 17).

The above does not imply that objectivist interpretations are false but it highlights why we might want to seek a subjective alternative. By "subjective" interpretations, I understand interpretations for which subjective/operational notions such as "measurement" or "experience" are central and irreducible. One such subjective interpretation is QBism. Recently, QBism has gained considerable attention, being called the "acme" of Copenhagen-style interpretations (Timpson 2013, 7) and the "most consistent and best-developed" (Berghofer & Wiltsche 2023) interpretation centered around the notion of experience.

The distinctive idea of QBism is to apply a personalist Bayesian account of probability, as developed by Bruno de Finetti, to quantum probabilities (Fuchs et al. 2014). Probabilities in quantum mechanics are interpreted as subjective probabilities and quantum states as representing an agent's subjective degrees of belief about her future experiences. That is, instead of being construed as (representing) something physically real, the wave function is considered to be a mathematical tool that encodes one's expectations about one's future experiences. In short, according to QBism quantum states do "not represent an element of physical reality but an agent's personal probability assignments, reflecting his subjective degrees of belief about the future content of his experience" (Fuchs & Schack 2015, 1). A measurement is understood as an act of the subject on the world and the outcome of a measurement is the very experience that results from this process (see DeBrota & Stacey 2019). Even probability-one predictions are considered subjective assignments expressing the agent's highest possible degree of certainty about what they will experience next. As Chris Fuchs puts it, a probability-one assignment "expresses the agent's supreme confidence that the outcome will occur" or "what the agent believes with his heart of hearts" (Fuchs 2023, 109). Here is an instructive quote:

> For instance, consider an agent who takes the action of placing a Stern-Gerlach device in front of an electron and has just registered spin-up for it in the z-direction as her consequent experience. She will thus assign a quantum state $|z = +1\rangle$ for any subsequent measurements on the electron. However, in QBism this does not amount to a statement of fact but a statement of belief. The assignment of this state amounts to, among other things, a belief—a monumentally strong belief—that taking the same action with the Stern-Gerlach device will give rise to exactly the same consequence, namely the experience of spin-up in the z-direction. (Fuchs 2023, 108)

For QBists, the quantum probabilities are degrees of belief. This means that for an event X, the probability P(X) represents the subject's degree of belief that X will occur. Crucially, QBists explicitly deny that quantum states or quantum probabilities represent what the subject *should* believe (personal communication). As we will see below, QBism does have an important normative dimension in the sense that quantum mechanics is viewed as a tool to help the agent determine whether her beliefs are *consistent*. But this is as far as it goes. Quantum states are purely subjective, quantum probabilities are purely subjective, and quantum mechanics is silent on what the subject should believe. The central issue addressed in this paper is that, if this is true, it becomes difficult to see how objectivity can enter science. I discuss this problem in the next section and propose a solution in Section 3.

**2. Is QBism too subjective?**

In my opinion, many objections against QBism articulating the idea that QBism is too subjective are unconvincing, and at least partly rest on misconceptions of QBism. In the following, I first discuss three common objections that I believe are based on misunderstandings; I then go on to discuss three objections that, in my view, pose serious problems. In Section 3, I introduce my degrees of epistemic justification interpretation (DEJI) and show how this interpretation avoids these problems.

First, QBism is often dismissed from the get-go for being "anti-realist." One of the earliest points of criticism in this direction is Hagar's assessment that the QBist approach constitutes an "instrumentalism in disguise" (Hagar 2003, 772). As Glick has correctly noted, "it remains commonplace among philosophers to take QBism to be opposed to scientific realism about quantum theory" (Glick 2021, 2). As also pointed out by Glick, this can be illustrated by QBism being classified as a "non-realist" approach by Myrvold in his *Stanford Encyclopedia* entry (Myrvold 2022). Typically, this assessment of QBism as anti-realist is intended as a strong objection to QBism (e.g. de Ronde 2016, 22). Importantly however, this non-realism must be qualified. QBism is non-realist regarding the wave function. But as mentioned above, there are good reasons to be skeptical about reifying the wave function and even some prominent proponents of objectivist interpretations oppose wave function realism. Here it is worth mentioning that QBists explicitly object to being labeled as anti-realists. Fuchs, in particular, insists that QBism must be understood "as being part of a realist program, i.e., as an attempt to say something about what the world is like, how it is put together, and what's the stuff of it" (Fuchs, 2017, 117). Regarding the external world, Fuchs says: "We believe in a world external to ourselves precisely because we find ourselves getting unpredictable kicks (from the world) all the time" (Fuchs 2017). Importantly, none of the QBist core claims implies any form of anti-realism regarding the external world (see also Glick 2021, 4).

Second, relatedly, QBism is often dismissed for not subscribing to a *ψ-ontic* interpretation of the wave function. Being not *ψ-ontic* distinguishes QBism from all of the objectivist interpretations. QBists reject *ψ-ontic* interpretations because QBists deny that the wave function represents an ontic state. Unfortunately, in the literature it is often mistakenly assumed that all interpretations that are not *ψ-ontic* are ruled out by the infamous PBR theorem (e.g. Maudlin 2019, 83-89). Here the thinking is that PBR (Pusey et al. 2012) rules out *ψ-epistemic* interpretations and that interpretations can only be *ψ-ontic* or *ψ-epistemic*. Importantly, however, QBists insist that quantum states neither represent an ontic state nor our knowledge/uncertainty of an underlying ontic state. QBism is more radical than that. QBism can be characterized as a *ψ-doxastic* interpretation since it holds that the quantum state represents subjective degrees of belief. PBR is silent on such a claim (DeBrota & Stacey 2019, Berghofer 2024a, Glick 2021, Hance et al. 2022) and thus this objection fails. Of course, you may still dismiss QBism for being non-representational. (Given the above, "non-representational" seems to be a better label for QBism than "non-realist.") But if you do, you do so for reasons of interpretational preference (you prefer representational interpretations over non-representational ones), not because any technical result such as the PBR theorem would rule out non-representational interpretations. Also, it is to be noted that even proponents of representational interpretations often point out that agent-centered approaches according to which quantum mechanics should be understood as a "single user theory" have the virtue of being capable of providing straightforward explanations of some of the weirdest features/implications of quantum mechanics (see, e.g., Dieks 2022). In particular, from the objectivist perspective it is, at least prima facie, disturbing that in certain scenarios, such as the Wigner's friend thought experiment, different agents assign different wave functions to the same physical state, use different types of dynamics to describe the same physical process, and that the collapse of the wave function occurs for the respective agents at different times.[2] From the QBist perspective (and agent-centered approaches in general), this is not surprising at all but is implied by quantum mechanics being understood as a single-user theory. Different agents will assign different wave functions to the same physical state if they have a different experiential input. The collapse occurs at different times if their update (new experiential input) happens at different times. This single-user aspect leads us to the next common objection.

Third, it is often argued that QBism leads to solipsism (Norsen 2016, Earman 2019, de Ronde 2016). If this objection is understood as saying that QBism *implies* solipsism, then this objection is clearly wrong. QBism says that the wave function only represents an agent's subjective degrees of belief about their future experiences, but that does *not* imply that there only is an agent and their experiences. In the QBist picture, quantum mechanics is a tool that can be used by an agent in order to improve their decision-making. In this context, the Born rule is viewed as a *normative* constraint that "functions as a consistency criterion which puts constraints on the agent's decision-theoretic beliefs" (Schack 2023, 146; see also DeBrota et al. 2021). This seems to imply that according to QBism quantum mechanics is silent on ontological matters such as the question of whether other subjects exist. But this does not imply that QBism implies solipsism.

Fourth, however, there is the worry that if QBism were true, it would be miraculous how objectivity or "objective knowledge" could enter science at all. This worry has the following structure. Quantum mechanics is our most fundamental scientific theory. If QBism is right, then quantum states are subjective assignments, quantum probabilities are subjective degrees of belief, and measurement outcomes are subjective experiences. But how, then, could the sciences lead to objective results?

This worry is particularly obvious if we view quantum probabilities as the output of the quantum machinery (Born rule). If both the input (quantum states) and the output (quantum probabilities) are purely subjective, then there is no room for objectivity to enter. Importantly, this was *precisely* the view specified by the QBists in (Fuchs et al. 2014). Here Fuchs, Mermin, and Schack emphasize that "the fundamental output of the quantum theory is not a set of facts, but a set of probabilities" (753). In this context, I want to highlight two points. First, I consider it natural and plausible to regard quantum probabilities as the output of the Born rule, and I assume that most physicists share this view. Second, some prominent QBists have since moved in a different direction. Fuchs and Schack now consider quantum probabilities as part of the input, with the output being simply "consistent" or "not consistent" (Schack 2023 and Fuchs 2023). This means that the role of the Born rule is to help the agent to be consistent in their decision-making. Regarding the question of how objectivity can enter science: I believe that QBists who regard subjective quantum probabilities as the fundamental output can only bite the bullet and assume that quantum mechanics cannot be linked to objective results. On the other hand, QBists who view the quantum formalism as fundamentally a decision-theoretic tool serving as a consistency criterion can link quantum mechanics to objective statements about consistency – but this is as far as it goes. There is nothing beyond consistency, not even on why consistency should matter.

---

2  For more details on the Wigner's friend thought experiment see (Fuchs 2023, Section 4) and (Baumann & Brukner 2023). For how the recently widely-discussed *extended* Wigner's friend scenarios support perspectvist interpretations, see (Schmid et al. 2023, 30).

In this paper, I aim to establish a much firmer link to objectivity and objectively justified beliefs by interpreting quantum probabilities as objective degrees of justification. In this picture, quantum mechanics is a machinery that has a subjective input (the wave function that is supposed to encode the agent's experiential input) but an *objective output* (quantum probabilities understood as degrees of epistemic justification).

Fifth, how could quantum mechanics be falsified if it only tells us something about subjective degrees of belief? In QBism, quantum probabilities are not associated with what you *should* believe but only represent your degrees of belief. In order to formulate the corresponding worry in a precise manner, let us consult how Dieks summarizes the QBist interpretation of quantum probabilities:

> Inherent in this doctrine is the interpretation of quantum probabilities as subjective. That is, QBist probabilities do not reflect relative frequencies, objective chances, or some other notion of physical probability; they rather serve to quantify personal, subjective degrees of belief. The subjective nature of QBist probabilities is illustrated by the meaning given to probability-1 statements. If a QBist agent predicts an experimental result with probability 1, this does not imply anything about the physical status of that future result; in particular, it does not entail that the result will necessarily be realized, or that the result in question is already there in the external world, waiting to be revealed. The only implication is that the agent is completely convinced that the outcome in question will be found. This is a fact about her or his expectations, not about the physical world. (Dieks 2022, 3f.)

I consider this an accurate and fair summary of the QBist stance. Now the problem can be formulated as follows: If the only implication is that the agent is convinced, and if quantum mechanics is only about expectations, then it is indeed difficult to see how quantum mechanics could be falsified. If quantum mechanics implies that an agent is convinced to experience A but then they experience non-A, this does not imply that there is anything wrong with quantum mechanics.

However, if quantum mechanics implies that the agent *should* expect to experience A, and then they experience non-A (e.g. in 90% of the relevant cases), this suggests that quantum mechanics suffers from shortcomings, is not (perfectly) reliable. According to the approach I present below, quantum mechanics can be compared in its epistemological function to a source of justification-conferring advice. Since quantum mechanics has been reliable in the past, and you have no evidence to distrust it, you should believe according to its advice. However, if its advice turned out to be in tension with your actual experiences, this would put pressure on its reliability and suggest that there might be a more reliable source of advice.

Sixth, it is not clear how anything like subjective degrees of belief could be quantified. Above, we have seen that QBists consider a probability-one prediction as a subjective assignment that expresses "the agent's supreme confidence," "what the agent believes with his heart of hearts," or "a monumentally strong belief." Does it mean that any agent making such an assignment actually undergoes a belief that has such a phenomenal quality? Does it mean that when different agents make the same assignment, they all have exactly the same belief? How to understand a situation in which an agent assigns probability 0.74 to outcome A? Does the agent actually believe that it is 0.74 likely that outcome A will occur or does she believe to degree 0.74 that outcome A will occur?[3] Neither seems plausible. Of course, this is not to say that QBists don't offer a consistent approach on these questions. Following the lead of Frank Ramsey, degrees of belief are quantified in terms of *betting behavior*: If a subject S assigns a probability of 0.74 to some event E, this means that when presented with a lottery ticket that offers a $1.00 payout if E happens and $0.00 otherwise, S would buy the lottery ticket for any amount less than $0.74 and sell it for any amount larger than $0.74. However, this betting interpretation of degrees of belief has been attacked on many fronts and it has been observed that "numerically sharp degrees of belief are psychologically unrealistic. It is rare, outside casinos, to find opinions that are anywhere near definite or univocal enough to admit of quantification" (Joyce 2010, 283). My point here is simply that objective degrees of epistemic justification can be more straightforwardly quantified than subjective beliefs.

## 3. The degrees of epistemic justification interpretation

My degrees of epistemic justification interpretation (DEJI) says the following: Take QBism as your role model but interpret quantum probabilities not as subjective degrees of belief but as *objective degrees of epistemic justification*. In terms of probability theory, this suggests that quantum probabilities should not be understood as subjective Bayesian probabilities along the lines of Bruno de Finetti but rather as objective Bayesian probabilities in the Coxian sense according to which probabilities represent *reasonable* expectations. This is to say that "the primary meaning of probability" is its function to provide a "measure of reasonable expectation" (Cox 1946, 2). In fact, what is now known as QBism started as a kind of objective Quantum Bayesianism. Here is how DeBrota and Stacey summarize why QBists moved from a Coxian understanding of probability to the one of Bruno the Finetti:

> Looking back on it, the attraction to the one over the other cuts to a rather fundamental point: QBism regards physics, and science in general, in Darwinian terms. The mathematics we develop is practical because, at root, it helps agents to survive. From this point of view, the idea of a probability as a gambling commitment, a belief made quantitative and ready to be acted upon, is an attractive notion. On the other hand, the idea of probability being used for a 'theory of inference' in the usual sense — i.e., a measure of plausibility for something that is 'out there' but unknown — is a bit off-putting. (DeBrota & Stacey 2019)

---
3 For a discussion of these two options on how to spell out the nature of credences, see (Moss 2018, 3f).

I agree with the QBists that the mathematics we use in our scientific theories is practical because it helps us to survive. But it seems reasonable to assume that it helps us to survive because our scientific theories possess some kind of epistemic virtue that can be linked to objectivity. Standard scientific realists may say that our scientific theories are objective in the sense that they are objectively true. But at least if you have some sympathies for perspectivist approaches to philosophy of science, this does not seem to be a useful way of putting it.[4] What I would insist on, however, is that our most successful scientific theories can tell us what we *should* believe, epistemically speaking. For instance, if quantum mechanics tells us that there is probability-one for outcome A, it is "objectively true" that we should believe we will experience outcome A. Importantly, however, I agree with the QBists that we should not go in the direction of hidden variables in the sense that quantum states represent something that is out there but unknown.

In this context, it is highly important to note that DEJI differs from the kind of objective Quantum Bayesianism that was proposed in the early days of Quantum Bayesianism. In particular, in 2002 Caves, Fuchs, and Schack introduced Quantum Bayesianism as an epistemic interpretation, explicitly arguing that "quantum states are states of knowledge" (Caves et al. 2002). More precisely, they said:

---

4   Here is how Hasok Chang criticizes this standard realist ideal: "Seeking absolute truth is not an operational ideal – there is nothing we can actually do in order to approach that ideal. According to the common picture of scientific knowledge, science should give us the true picture of the reality that exists well-formed 'out there' completely independently of our conceptions and our experiences. But such 'reality' is not accessible to us and there are no actual methods by which we could attain assured knowledge about it." (Chang 2022, 2)

> We accept the conclusion of Einstein's argument and start from the premise that 'quantum states are states of knowledge.' An immediate consequence of this premise is that all the probabilities derived from a quantum state, even a pure quantum state, depend on a state of knowledge […] If two scientists have different states of knowledge about a system, they will assign different quantum states, and hence they will assign different probabilities to the outcomes of some measurements. (Caves et al. 2002)

Here Caves et al. basically subscribe to what is now known as a *ψ-epistemic* interpretation of the wave function. One reason why I do not call my interpretation simply "Objective Quantum Bayesianism" is that I reject this idea that quantum states are states of knowledge about an underlying ontic state. I am sympathetic to the QBist (Bohrian) move to focus on the notion of experience and to consider the wave function an assignment of the agent that encodes the agent's experiences. In my interpretation, the wave function is an assignment and the quantum probabilities tell the agent what she *should* believe to experience next based on this assignment. So there is an agent in the world who has some experiential input and quantum mechanics tells the agent what, based on this experiential input, she should believe to experience next. Central and irreducibly primitive are the concepts of agent (subject), experience, and justification. DEJI differs from the objective Bayesianism of Caves et al. 2002 in that the quantum state does not represent knowledge but is an assignment of the agent based on her experiences – the concept of experience plays no role in Caves et al. 2002 – and in that the notion of knowledge is not treated as fundamental in my account. DEJI differs from QBism in that quantum probabilities are not subjective degrees of belief but objective degrees of justification.

In Fuchs 2002, Fuchs captures the development of his thinking about quantum states as follows: "*Knowledge → Information → Belief → Pragmatic Commitment*." For an agent- or experience-centered approach, the notions of knowledge and information are problematic because they are *factive* concepts. If I know (have the information) that *p*, then *p* is the case. The notions of belief and pragmatic commitment, in my view, are too subjective if this is to say that the outcomes of an agent's usage of quantum mechanics, namely the quantum probabilities, are understood simply as the agent's subjective degrees of belief or pragmatic commitment. Instead, I view quantum mechanics as a machinery into which we feed some experiential input (encoded in the wave function), and as an output, we get the answer to the question of what we *should* believe to experience next, based on this experiential input. This output comes in form of objective degrees of epistemic justification.

In short, DEJI is an agent-centered interpretation that views quantum mechanics as a single-user theory that allows an experiencing subject to answer the following question: Based on my experiential input, what should I believe to experience next? The input is the wave function and the output is quantum probabilities. The wave function is an assignment of the agent that is supposed to encode the agent's experiential contact with the world. The quantum probabilities are degrees of epistemic justification that encode – based on the experiential input – what the agent should believe to experience next. In DEJI (as well as in QBism) different agents can ascribe different wave functions to the same physical situation due to the simple fact that different agents can have a different experiential input. As in Wigner's friend scenario, this can happen if the agents are physically separated, having experiential access to different parts of the world. However, this can also happen in examples in which different agents have experiential access to the exactly same parts of reality, for instance, if one agent assigns their wave function based on a measurement error.[5]

DEJI shares many systematic similarities with Healey's pragmatist approach to quantum theory. DEJI and Healey's pragmatism are both non-representational interpretations according to which quantum mechanics does not describe the behavior of an external reality but has a crucial normative dimension in telling the agent what they should believe/do. They both consider the agent-centeredness of QBism a virtue but seek to establish an interpretation that is more objective than QBism. For Healey, quantum theory "is a source of objectively good advice about *how* to describe the world and what to believe about it as so described. This advice is tailored to meet the needs of physically situated, and hence informationally-deprived, agents like us" (Healey 2022, Section 4.3). The main differences between DEJI and Healey's pragmatism are that in the latter there is no focus on the notion of experience and no talk in terms of epistemic justification. It is to be noted that when it comes to epistemic justification, I'm not a pragmatist but champion the following "objective" view: The degree of propositional justification a subject has for believing some proposition *p* is an objective matter of fact that is independent of the subject's goals, wishes, or desires.[6]

**4. Virtues and challenges**

The idea is that DEJI has all the virtues of QBism but, in contrast to QBism, allows objectivity to enter science. By "all the advantages of QBism" I mean that as a proponent of DEJI you neither need to modify the formalism nor accept many worlds. You do not need to worry about how to make sense of the wave function ontologically or how to understand the collapse of the wave function physically. You can accept the Bohrian intuition that an interpretation of quantum mechanics should be centered around the notion of experience, without invoking the notion of complementarity. Furthermore, you are not committed to non-local dynamics (spooky action at a distance), and you can provide a straightforward approach to the Wigner's friend scenario. All this is achieved by insisting that the wave function is not a physical object and that the collapse of the wave function, accordingly, is not a physical process but an update connected to what the agent experiences. In Wigner's friend scenario, it is not surprising that Wigner and his friend assign different wave functions and that Wigner's wave function collapses later than the one of his friend. This is merely a consequence of the fact that they have a different experiential input and that Wigner's experiential update happens later than the one of his friend. The Born rule is not something that is required to be derived as in the many-worlds interpretation but constitutes what is essential and fundamental about quantum mechanics: It allows you to connect your experiential input with objective degrees of epistemic justification. This means that objectivity enters quantum mechanics from the get-go. Also, it is clear how quantum mechanics can be falsified. If your experimental results do not match with what, according to quantum mechanics, you should have expected to experience, then this suggests that quantum mechanics is not perfectly reliable in ascribing degrees of epistemic justification to the contents of possible future experiences.

---

5     Since the quantum state has the function to encode what the subject experiences, and since experiences represent the experienced world, quantum states represent the world in an indirect sense.
6     One might wonder how subjective experience can lead to objective justification. Importantly, epistemology has a rich history of treating experiences as justifiers and there is a huge literature on how experiences justify. I introduce and discuss some epistemological distinctions and positions that are relevant in this context in the appendix. Philosophically, DEJI is best understood within the framework of an experience-first epistemology as elaborated in the appendix.

**List of Virtues**

In a nutshell, if you are a proponent of DEJI, you

V1: don't need to modify quantum mechanics (in contrast to, e.g., Bohmian mechanics and GRW),

V2: don't need to accept many worlds (in contrast to MWI),

V3: don't need to derive the Born rule but can treat it as a primitive (in contrast to, e.g., MWI),

V4: don't need to accept non-local dynamics (in contrast to, e.g., Bohmian mechanics and GRW),

V5: don't need to worry about how to make sense of the wave function ontologically (in contrast to all objectivist interpretations and some versions of the Copenhagen interpretation that are unclear on the nature of the wave function),

V6: don't need to worry about how to make sense of the collapse of the wave function physically (in contrast to some versions of the Copenhagen interpretation that consider the collapse a physical process),

V7: don't need to center your understanding of quantum mechanics around the notion of complementarity (in contrast to Bohr),

V8: don't need to operate with factive concepts such as "knowledge" that don't resonate well with an agent-centered interpretation (in contrast to, e.g., the objective Bayesianism of Caves et al. 2002),

V9: don't need to commit to a *ψ-epistemic* interpretation of the wave function that may be ruled out by the PBR theorem (in contrast to interpretations that are centered around the notion of "knowledge"),

V10: don't need to worry about how objectivity could enter science (in contrast to QBism),

V11: don't render quantum mechanics an unfalsifiable theory (in contrast to QBism),

V12: can straightforwardly make sense of the Wigner's friend scenario,

V13: advocate an interpretation of quantum mechanics that seems to lead to a straightforward connection between science and epistemology (as explained in the appendix).

Even if you are a strong proponent of objectivist interpretations, you should agree that each feature of V1-V12 can be considered a virtue. (I assume most will be neutral on V13, which will be discussed in the appendix.) If you are a Bohmian, for instance, you may think that proposing a clear ontology is more important than V1, but you should not think that V1 is not a virtue.

This brings me to the main challenge of DEJI. This is a challenge that also QBism struggles with, namely the challenge to answer the following questions. What does quantum mechanics tell us about the world? What are the ontological implications? Think about classical mechanics. In classical mechanics, interpreting physics seems easy. You have point particles that "live" in three-dimensional Euclidean space. These particles have a definite position and momentum. And your equations of motion determine the evolution of these particles. Here it is natural to assume that the point particles and their positions and momenta can be understood as representing physical objects and real physical properties. If DEJI (or QBism) is true, however, quantum mechanics does not offer this clear ontology we know from classical mechanics. Its basic object, the wave function, does not represent a physical object or state. The evolution of this "object" cannot be solely described from a deterministic, third-person perspective (Schrödinger dynamics), but this dynamics breaks down when a measurement happens. This collapse of the wave function is linked to the experiential update of the agent. In other words, if DEJI (or QBism) is true, then we need to rethink the nature and purpose of science. At a fundamental level, then, science does not provide us with a third-person description of how external objects evolve in time, but tells us what we should believe to experience next based on our experiential input. So what should we reply to the proponent of objectivist interpretations who clearly favors an understanding of science as motivated by classical mechanics over the one delivered by DEJI?

First, it seems reasonable to propose that our interpretation/understanding of science should be based on our *actually* most successful scientific theories/models. It should *not* be based on theories that have been outdated for over a century. Thus, instead of tampering with quantum mechanics until it fits our ontological intuitions and our understanding of science we both inherited from classical mechanics, we should approach quantum mechanics as open-mindedly as possible. It has been argued that the success story of modern physics can be constructed as a story of abandoning metaphysical hypotheses (Mittelstaedt 2011) and that physics has been developing toward theories that "do away with the idea of entities" (Grinbaum 2017). Of course, this is not to say that DEJI is prima facie more plausible than, for instance, Bohmian mechanics. However, the current situation in philosophy of quantum mechanics is that more philosophers are working on modificatory interpretations than on trying to make philosophical sense of subjective interpretations. The latter endeavor has not however been pursued primarily by philosophers, but rather by physicists such as Bohr and Heisenberg. The modest proposal here is that more philosophers should be working on this.

Second, not even classical mechanics can be interpreted/understood as straightforwardly as it is often assumed. This is because also in classical mechanics the same physics can be formulated in different mathematical frameworks. Newtonian, Lagrangian, and Hamiltonian mechanics constitute different mathematical frameworks for solving the same physical problems. Is classical mechanics sufficiently clear such that any further questions of interpretation become superfluous? The mere existence of philosophical discussions in classical mechanics suggests that they are not (see North 2022 and Wilson 2013). Let's consider a classical system of N particles. Typically we say that here the "state of the system is represented by $N$ points $X_1,...X_N$, in three-dimensional Euclidean space" (Wallace 2021, 68). Importantly, this is only one possible mathematical representation. As Wallace points out "there is another way to represent this theory. We can define the configuration space as the product of $N$ copies of Euclidean three-space. Each $N$-tuple of points $(X_1,...X_N)$ now corresponds to a single point in this $3N$-dimensional space" (Wallace 2021, 68). Wallace discusses this in the context of his objections against wave function realism, arguing that it would be misleading to consider this $3N$-dimensional configuration space as the true space we live in. I agree. I would add that this also illustrates how problematic it can be to take the mathematical representation of some theory at face value, trying to read off an ontology directly. This brings us to the quantum reconstruction program discussed next.

Third, in my view, the most consistent answer that proponents of DEJI and QBism can give to the objection that DEJI/QBism does not provide us with a clear ontology is to insist that before we ontologically interpret a highly mathematical physical theory, we should first have a conceptual grasp on the question of where the mathematics comes from. This is precisely the objective of the so-called *quantum reconstruction program* that recently gained some prominence among physicists working on the foundations of quantum mechanics as it turned out that the quantum formalism can be derived from operationally meaningful information-theoretic principles.[7] The cornerstones of this program have been clarified as follows:

> In short, the postulates of quantum theory impose mathematical structures without providing any simple reason for this choice: the mathematics of Hilbert spaces is adopted as a magic blackbox that 'works well' at producing experimental predictions. However, in a satisfactory axiomatization of a physical theory the mathematical structures should emerge as a consequence of postulates that have a direct physical interpretation. By this we mean postulates referring, e.g., to primitive notions like physical system, measurement, or process, rather than notions like, e.g., Hilbert space, $C^*$-algebra, unit vector, or self-adjoint operator. (D'Ariano et al. 2017, 1)

This program rests on the idea that we can improve our understanding of quantum mechanics by deriving (reconstructing) the formalism from meaningful physical principles. The role model here is special relativity (see, e.g., Rovelli 1996, 1639; Zeilinger, 1999; Chiribella & Spekkens 2016, 3). As is well known, Einstein did not discover the mathematical formalism of special relativity which was already present in Lorentz' work several years before Einstein's annus mirabilis. However, what Einstein achieved was to show how this formalism follows from two meaningful principles, namely the special principle of relativity and the light postulate. Our understanding of special relativity is based on more than a mathematical understanding of the formalism. We understand special relativity because we understand the foundational principles it relies on. This understanding has been improved by further "reconstructions" such as Minkowski's insight that, alternatively, the formalism of special relativity can also be derived from the single postulate of Minkowski spacetime. As of yet, there exist several successful reconstructions of the quantum formalism, e.g., Hardy 2001, Chiribella et al. 2011, Dakic & Brukner 2011, Masanes et al. 2013, Goyal 2014, Höhn 2017, and Höhn & Wever 2017. Although, unfortunately, so far these reconstructions have not revolutionized our understanding of the quantum formalism, all successful reconstructions have something in common: They are formulated in an operational framework and are based on *information*-theoretic principles. There is no indication that reconstructions could also be achieved within an objectivist framework, e.g., taking a particle ontology or many-worlds ontology as a starting point (see Koberinski & Müller 2018, 265 and Fuchs 2014, 388). This is not to say that the quantum reconstruction program rules out objectivist interpretations but I would say it is a virtue of an interpretation of quantum mechanics if it resonates well with successful reconstructions (e.g., Bub 2018). Returning to our question of how to respond to the objection that DEJI/QBism does not have a clear ontology, I side with Fuchs' sentiment that "it became clear that the pertinent way to move forward was to get the 'epistemics' of the theory right before anything else: Getting reality right would follow for those who had patience enough to pass the marshmallow test" (Fuchs 2023, 78).[8]

---

7  For philosophical reflections on the quantum reconstruction program, see, e.g., Adlam 2022, Berghofer 2024b, and Grinbaum 2006, 2007.

**Conclusion**

If you neither want to modify the formalism of the most successful scientific theory in the history of science nor accept infinitely many worlds that are, in principle, unobservable, then you should be interested in subjective interpretations of quantum mechanics, i.e., interpretations in which subjective/operational terms such as "experience" or "measurement" play a central role. QBism is probably the currently best-developed interpretation that is centered around the notion of experience. For many, however, QBism is too subjective. According to QBism, quantum states and quantum probabilities are subjective. But if this is true, how could objectivity enter science? How could quantum mechanics be falsified? What is more, it is unclear whether degrees of subjective belief can occur or be quantified in a manner as fine-grained as quantum probabilities. In this paper, I seek to resolve these problems by interpreting quantum probabilities not as subjective degrees of belief but as objective degrees of epistemic justification. If my calculations yield that outcome A is associated with a probability of 0.74, this means that my degree of justification for believing that I will observe outcome A is 0.74. As explained in the appendix, this justification is to be understood as propositional justification as opposed to doxastic justification, and thus it does not matter what I actually believe. Quantum probabilities are *objective* degrees of epistemic justification. Quantum mechanics tells you what you *should* believe.

---

[8] See in this context also Markus Müller's "first-person-first" approach to science (Müller 2020) that also seeks to characterize quantum probabilities in a more objective way than it is done by the QBists. I take it my epistemological account specified in this paper coheres well with Müller's technical results (see also Jones & Müller 2024, 21).

**Appendix: Epistemological clarifications**

**A1: Knowledge vs. justification**

One crucial difference between knowledge and justification is that knowledge is factive, but justification is not. Whatever you believe, whatever your evidence is, you cannot know that Einstein invented classical mechanics because it is not true that Einstein invented classical mechanics. If one knows that *p*, then *p* is the case. If you are undergoing a perceptual experience that presents you with a tree in front of you, but this experience is a non-veridical hallucination caused by some drug, then you might believe that there is a tree but you do not know it. You do not know it because it is not true. Justification, on the other hand, is non-factive. It is reasonable to assume that cases are possible in which you are justified in believing that *p*, even if *p* is not the case. Assume you are interested in a mathematical theorem T, but you do not know whether T is true. So you ask your friend, whom you know to be a capable mathematician and reliable source of information. Your friend assures you that T is true. Just to be sure, you also consult Wikipedia, and Wikipedia says T is true. Now let us stipulate that for whatever reason T is false. It is natural to assume that although (unbeknownst to you) T is false, you are justified in believing that T. Returning to our example of hallucination, if you are undergoing a perfect hallucination of a tree, and you have no evidence for believing that this experience is hallucinatory, it is reasonable to believe that you are justified in believing that there is a tree.

Let us now return to the objective Quantum Bayesianism of Caves et al. 2002 according to which "quantum states are states of knowledge" such that "[i]f two scientists have different states of knowledge about a system, they will assign different quantum states, and hence they will assign different probabilities to the outcomes of some measurements." Since knowledge is factive, this means that the two scientists make different assignments to the same system that are both true. This is already a bit strange. But the deeper problem is why need the assignment of an agent be veridical in the first place? If your approach to quantum mechanics is an agent-centered one in which the agent makes assignments according to what they experience, this does not make much sense as pointed out by the QBists (Fuchs et al. 2014, 753): "Anybody using quantum mechanics to organize her experience can be an agent, and different agents have different experiences." They continue by pointing out: "Second, 'knowledge' is the wrong word because the fundamental output of the quantum theory is not a set of facts, but a set of probabilities" (Fuchs et al. 2014, 753). As discussed above, I agree that the fundamental output of quantum theory is a set of probabilities and that this is not adequately captured in terms of knowledge. It is not, because it is not natural to think of knowledge in terms of (fine-grained) degrees. To say that my degree of knowledge regarding the proposition that "I will observe outcome A" is 0.57 is a very strange statement. The QBists are right that beliefs come in degrees. But it is also very strange to say that my degree of belief regarding the proposition that "I will observe outcome A" is 0.57. Beliefs are not that fine-grained. However, justification also comes in degrees and if we consider quantum mechanics as a formalism that allows me to quantify what I *should* believe, it is not a surprise that objective degrees of epistemic justification come in such a fine-grained shape.

**A2: Propositional vs. doxastic justification (justification vs. belief)**

In epistemology, it is common to distinguish between propositional and doxastic justification. Basically, this corresponds to the distinction between "*having justification to believe that p* versus *justifiedly believing that p*" (Silva & Oliveira forthcoming). The former characterizes propositional justification, denoting whether a subject has justification to believe some proposition (whether or not the subject actually believes the respective proposition); the latter characterizes doxastic justification, denoting whether a subject's actual belief is justified. Imagine you read on Wikipedia that Napoleon was born in 1769. You know Wikipedia to be a reliable source on such questions and thereby have propositional justification to believe that Napoleon was born in 1769. However, let us also assume that for whatever (irrational) reasons you do not actually believe it. This does not change the fact that you have propositional justification for believing it. Propositional justification is always justification of some subject, but this subject can have propositional justification without having any corresponding belief. This is what makes propositional justification so well-suited for my approach. Given some experiential input, quantum mechanics tells you what you should believe, whether or not you actually believe it. Epistemic justification and quantum mechanics are both subjective in some sense and objective in another sense. Justification is subjective because it is always a subject that is justified. This distinguishes justification from truth. The proposition that "Napoleon was born in 1769" is either true or false. It does not depend on any subject whether the proposition is true. However, this proposition is not either justified or not justified. *Believing* this proposition is either justified or not justified *for some subject*. Justification is objective because a subject is justified to believe some proposition whether or not the person actually has any epistemic attitude toward this proposition.[9] If a subject has sufficient evidence (reasons) to believe that *p*, this subject is justified to believe that *p* whether or not the subject believes, doubts, or disregards *p*.

---

9   Accordingly, it is important to distinguish between epistemic justification and epistemic attitudes. For instance, when I have 0,5 justification for believing *p*, this does not mean that I am epistemically justified in *believing* that *p*. In other words, the adequate epistemic attitude in this case is not belief but suspension of judgment.

Quantum mechanics is subjective as well as objective because it tells some *subject* what it *should* believe. Quantum mechanics does not constitute a third-person description of how external objects evolve in time. Instead, it can be viewed as a machinery which you can provide with some subjective input, and as an output you get degrees of propositional epistemic justification. This output tells you what – given your experiential input – you should believe to experience next. This output is objective because the degrees of propositional epistemic justification apply notwithstanding what you actually believe.

**A3: From epistemology to science**

A substantial part of contemporary analytic epistemology has been occupied with the task of precisely defining the concept of knowledge. The failure to find a definition that enjoys widespread agreement has led several philosophers to believe that it might be more fruitful to center epistemology around notions that are more easily accessible. One such proposal is that epistemology should be centered around the notion of experience (see, e.g., Dougherty & Rysiew 2014, Berghofer 2022). In this spirit, we say that experiences are our most basic justifiers, i.e., all epistemic justification, and every piece of knowledge, can be traced back to epistemically foundational experiences. This sentiment has been expressed in various ways. For instance: "Experience is our point of interaction with the world – conscious awareness is how we gain whatever evidence we have" (Conee & Feldman, 2008, 87). Let us assume that experience-first epistemologies are on the right track. This would mean that the central task of epistemology is to clarify the epistemic role of experience. This amounts to an a priori analysis of how experience relates to the justification of belief. But of course epistemology is silent on what I actually experience and on what I should believe to experience next. If DEJI is on the right track, science can be understood as filling this gap. Science provides us with the formalism that allows the experiencing subject to quantify what – based on her experiential input – she should believe to experience next. This formalism, of course, is the formalism of quantum mechanics. This would mean that there is an intimate relationship between epistemology and science, both being centered around the notion of experience such that we might think of science as a continuation of epistemology, or perhaps even as *applied epistemology*. This does not constitute an argument for favoring DEJI over rival objectivist interpretations of quantum mechanics, but it is an invitation for philosophers to contemplate the relationship between epistemology and science from an agent- and experience-centered perspective.